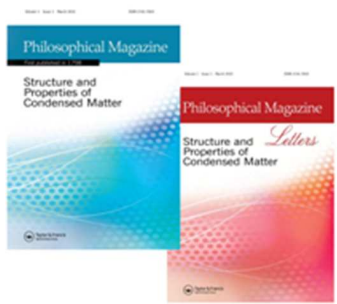

# An original unified approach for the description of phase transformations in steel during cooling

# An original unified approach for the description of phase transformations in steel during cooling

O. Bouaziz[*,1,2]

[1]ArcelorMittal Research, Voie Romaine-BP30320, 57283 Maizières-lès-Metz Cedex, France
[2]Centre des Matériaux, Ecole des Mines de Paris, CNRS UMR 7633, B.P. 87, 91003 Evry Cedex, France

Email : olivier.bouaziz@arcelormittal.com

## Abstract

Exploiting Landau's theory of phase transformations, defining an original order parameter and using the phenomenological transformation temperatures, it is reported that it is possible to describe in a global approach the conditions for the formation of each phase (ferrite, bainite, martensite) from austenite during cooling in steel. It allowed to propose a new rigorous classification of the different thermodynamic conditions controlling each phase transformation. In a second step, the approach predicts naturally the effect of cooling rate on the bainite start temperature. Finally perspectives are assessed to extend the approach in order to take into account the effect of an external field such as applied stress.

## 1. Introduction

In the field of solid-state phase transformation in metallic alloys [1], the transformation of austenite in steel on cooling can occur by a variety of mechanisms including the formation of ferrite, bainite and martensite. The bainitic transformation occurs in a range between purely diffusional transformation to ferrite or pearlite and low temperature transformation to martensite by a displacive mechanism. Thus the bainite transformation exhibits features of both diffusional and displacive transformations and has given rise to a large amount of research activity (see [2-3] for reviews). A major part of the research has concerned modelling of the kinetics of the transformations by detailed descriptions of the thermodynamic conditions operating at the interface [4-7]. However the crucial understanding of the physically based conditions of the start of each phase transformation is less understood, especially for bainite.

Usually the $Ar_3$ temperature is defined as the minimum temperature for any phase transformation of austenite to ferrite during cooling [8] :

$$Ar_3(^\circ C) = 910 - 230C - 21Mn - 15Ni + 45Si + 32Mo \qquad (1)$$

For temperature lower than $Ar_3$, the basical tools of physical metallurgy are the definition of the martensite start temperature $M_s$ and the bainite start temperature $B_s$ expressed phenomenologically as functions of chemical composition as :

$$M_s(^oC) = 539 - 423C - 30.4Mn - 17.7Ni - 12.1.Cr - 11Si \quad (2)$$

where alloying element are expressed in weigth% [9], and [10] :

$$B_s(^oC) = 870 - 270C - 90Mn - 37Ni - 70Cr - 83Mo \quad (3)$$

In addition another characteristic temperature is defined as the temperature where austenite and ferrite has the same thermo-chemical free energy determined as [2,11]:

$$T_o(^oC) = 835 - 198.C - 91Mn - 36Ni - 73Cr - 15Si - 87Mo \quad (4)$$

In this publication, it is showed that it is possible to describe in a global approach the conditions for each phase transformation exploiting completely Landau's theory [12-13] of phase transformations including an original parameter of order and to propose a new classification of the different phase transformation in steel during cooling and to predict naturally the effect of cooling rate on bainite start temperature.

## 2. The proposed approach

In the framework of Landau's theory [12-13], for a second order phase transition, the free energy is expressed as :

$$F(\chi, T) = F_o(T) + A(T)\chi^2 + C.\chi^4 \quad (5)$$

where T is the temperature, $\chi$ the order parameter, $F_o(T)$ the thermo-chemical free energy, $A(T)$ a function of temperature and C a positive constant.

The simplest expression for $A(T)$ is :

$$A(T) = A_o(T - T_c) \quad (6)$$

with $A_o$ a positive constant and $T_c$ a critical temperature where $A(T)$ changes of sign.

Usually in phase transformation of steels, $F_o(T)$ is known [2,3,7]. So the identification of the total free energy $F(\chi, T)$ requires the determination of three parameters : $A_o, T_c, C$.

For $T > T_c$, $F(\chi, T)$ exhibits one minimum as a function of $\chi$ :

$$\frac{\partial F(\chi, T)}{\partial \chi} = 0 \quad (7)$$

for $\chi = 0$

For $T \leq T_c$, $F(\chi, T)$ exhibits two minima as a function of $\chi$ :

$$\frac{\partial F(\chi, T)}{\partial \chi} = 0 \tag{8}$$

for

$$\chi = \pm \sqrt{\frac{A_o(T_c - T)}{2.C}} \tag{9}$$

In order to exploit this approach to a classification of phase transformations in steel, it is now assumed that :

$$T_c = B_s \tag{10}$$

where $B_s$ is the bainite start temperature.

By convention, it is chosen to have an order of parameter for martensite :

$$\chi = \sqrt{\frac{A_o(B_s - M_s)}{2.C}} = 1 \tag{11}$$

Giving a first relationship :

$$\frac{A_o}{2.C} = \frac{1}{B_s - M_s} \tag{12}$$

Therefore the order parameter is :

$$\chi = \sqrt{\frac{B_s - T}{B_s - M_s}} \tag{13}$$

In order to illustrate quantitatively this law, the evolution of the order parameter from Bs to Ms temperatures for two different Fe-C compositions has been plotted in Fig1.

It is now proposed to define clearly what could be the parameter of order for phase transformation in steels. If C is the carbon content of the phase appearing, it is reasonable to propose that :

$$\chi = \frac{C - C_{\alpha,eq}}{C_\gamma - C_{\alpha,eq}} \tag{14}$$

Where $C_{\alpha,eq}$ is the solubility of carbon in ferrite at equilibrium and $C_\gamma$ is the carbon in austenite. So the order parameter can be a sursaturation in carbon in the binary Fe-C system.

Figure 1. : Evolution of the order parameter from Bs to Ms temperatures for two different Fe-C compositions.

In addition chemical free energy at $M_s$ has been determined as [12]:

$$F_o(M_s, T) = F_o(M_s) + S_o.(T - M_s) \tag{15}$$

with $F_o(M_s) = 1250 J/mol$ and $S_o = -6.8 J.mol^{-1}.K^{-1}$, wich are independent of the composition.

The energy value $F_o(M_s)$ should correspond to the maximum at $M_s$ for $\chi = 0$. As for any $T < B_s$ this maximum exist for the same value of $\chi = 0$, it is written for $M_s \leq T < B_s$ :

$$F(0,T) = F_o(T - M_s) + F_o(M_s) \qquad (16)$$

with

$$F_o(T - M_s) = S_o.(T - M_s) \qquad (17)$$

when T is near $M_s$ but it can be completely different for higher temperature.

In order to determine $A_o$ and $C$, as the total free energy at $M_s$ for $\chi = 1$ has to be zero, it comes :

$$F_o(M_s) + A_o(M_s - B_s) + C = 0 \qquad (18)$$

or

$$C = 2C - F_o(M_s) \qquad (19)$$

and

$$C = F_o(M_s) \qquad (20)$$

Consequently :

$$A_o = \frac{2.F_o(M_s)}{B_s - M_s}. \qquad (21)$$

Finally it is possible to express completely the free energy :

$$F(\chi,T) = F_o(T - M_s) + F_o(M_s) + \frac{2.F_o(M_s)}{B_s - M_s}(T - B_s)\chi^2 + F_o(M_s)\chi^4 \qquad (22)$$

or

$$F(\chi,T) = F_o(T - M_s) - F_o(M_s) + F_o(M_s).\chi^2\left(2.\frac{(T - B_s)}{B_s - M_s} + \chi^2\right) \qquad (23)$$

In order to highlight the key role of the order parameter, it has been drawn in Fig3. the evolution of the right-end side term of the expression :

$$F(\chi,T) - F_o(T - M_s) = F_o(M_s) + F_o(M_s).\chi^2\left(2.\frac{(T - B_s)}{B_s - M_s} + \chi^2\right) \qquad (24)$$

Figure 2. : Evolution of the $F(\chi,T) - F_o(T - M_s)$ as a function of the order parameter at $B_s$ and at $M_s$ temperatures.

Finally, the quantitative developed approach can be used in order to summarized the thermodynamic conditions for the formation of each phase, as it is summarized in Table 1, providing more rigorous occurrence criterion especially to distinguish ferrite, bainite or martensite formation.

Tab.1 : Classification of phase transformation conditions from austenite during cooling

## 2. Extended approach including cooling rate

It is well known that the formation of the bainitic phase depends on the cooling rate applied to austenite [2]. It means that the critical temperature $T_c$ defined in Eq.6 can be chosen equal to $B_s$ only for the minimum cooling rate $C_{r,min}$ for bainite formation. A maximum cooling rate $C_{r,max}$ is also commonly used indicating the end of bainite formation and the start of martensite occurrence [2,16]. Therefore, in the case of a continuous cooling (i.e. constant cooling rate), it is also possible to define the order parameter as a function of cooling rate as :

$$\chi = \left(\frac{C_r - C_{r,min}}{C_{r,max} - C_{r,min}}\right)^p \tag{25}$$

where $C_r$ is the constant applied cooling rate and p is a parameter.

In order to have a complete consistency with the definition of $\chi$ determined in Eq13,

$$\sqrt{\frac{B_s - T}{B_s - M_s}} = \left(\frac{C_r - C_{r,min}}{C_{r,max} - C_{r,min}}\right)^p \tag{26}$$

The temperature T respecting Eq.26 is the "generalized" Bainite Start temperature including the effect of cooling rate expressed as :

$$B_{s,g} = B_s - (B_s - M_s)\left(\frac{C_r - C_{r,min}}{C_{r,max} - C_{r,min}}\right)^{2.p} \tag{27}$$

A value of the parameter p has been identied by comparison with experimental data of the evolution of $B_{s,g}$ as a function of cooling rate which can be extracted from Continuous Cooling Transformation diagram [17]. For the first assessment of the approach, a chemical composition as simple as possible has been selected (Fe-0.4%C-0.7%Mn-0.2%Si). The experimental evolution of $B_{s,g}$ is reported in Fig.3 and compared with the model (Eq.27). A very good agreement is found for p equal to $1/4$. It means that a law for $B_{s,g}$ can be expressed as :

$$B_{s,g} = B_s - (B_s - M_s)\left(\frac{C_r - C_{r,min}}{C_{r,max} - C_{r,min}}\right)^{1/2} \tag{28}$$

Figure 3. : Comparison between experimental and predicted evolution of the Bainite Start temperature as a function of cooling rate for Fe-0.4%C-0.7%Mn-0.2%Si composition.

## 4. Conclusions

Exploiting Landau's theory of phase transformations, defining an original order parameter and using the phenomenological transformation temperatures, it is reported that it is possible to describe in a global approach the conditions for the formation of each phase (ferrite, bainite, martensite) from austenite during cooling in steel. It allowed to propose a new more rigorous classification of the different thermodynamic conditions controlling each phase transformation. In a second step, the approach predicts naturally the effect of cooling rate on the bainite start temperature.

In perspectives, the approach can be extended to take into account external fields by adding a linear term in free energy linearly proportional to the order of parameter and proportional to the potential energy of the external field [12-13]. For instance, in the case of an uniaxial applied stress $\sigma$, the contribution to the free energy is expressed as :

$$F(\chi,\sigma)) = \pm \frac{\sigma^2}{2.E}.\chi \tag{29}$$

where E is the elastic modulus and the sign depends on the compressive or tensile stress.

This term breaks the symmetry of the total free energy as a function of the parameter of order and it can change the occurrence conditions of each phase. A lot of experimental data are available is the literature in order to validate this last point which will be investigated in a next future.

Acknowmedgement : The author thanks Dr. F. Levy for stimulating discussions.

**Tables with captions**

| Phase | Free energy | Order parameter |
|---|---|---|
| Ferrite | $\dfrac{\partial F(\chi, T)}{\partial \chi} = 0$ | $\chi = 0$ |
| Bainite | $\dfrac{\partial F(\chi, B_s)}{d\chi} = 0$ <br> $\dfrac{\partial^2 F(\chi, B_s)}{\partial \chi^2} = 0$ | $T < B_s, \chi = \sqrt{\dfrac{B_s - T}{B_s - M_s}}$ |
| Martensite | $F(\chi, M_s) = 0$ <br> $\dfrac{\partial F(\chi, M_s)}{\partial \chi} = 0$ | $\chi = 1$ |

Tab.1 : Classification of phase transformation conditions from austenite during cooling

**Figures captions**

Figure 1. : Evolution of the order parameter from Bs to Ms temperatures for two different Fe-C compositions.

Figure 2. : Evolution of the $F(\chi,T) - F_o(T - M_s)$ as a function of the order parameter at $B_s$ and at $M_s$ temperatures.

Figure 3. : Comparison between experimental and predicted evolution of the Bainite Start temperature as a function of cooling rate for Fe-0.4%C-0.7%Mn-0.2%Si composition.

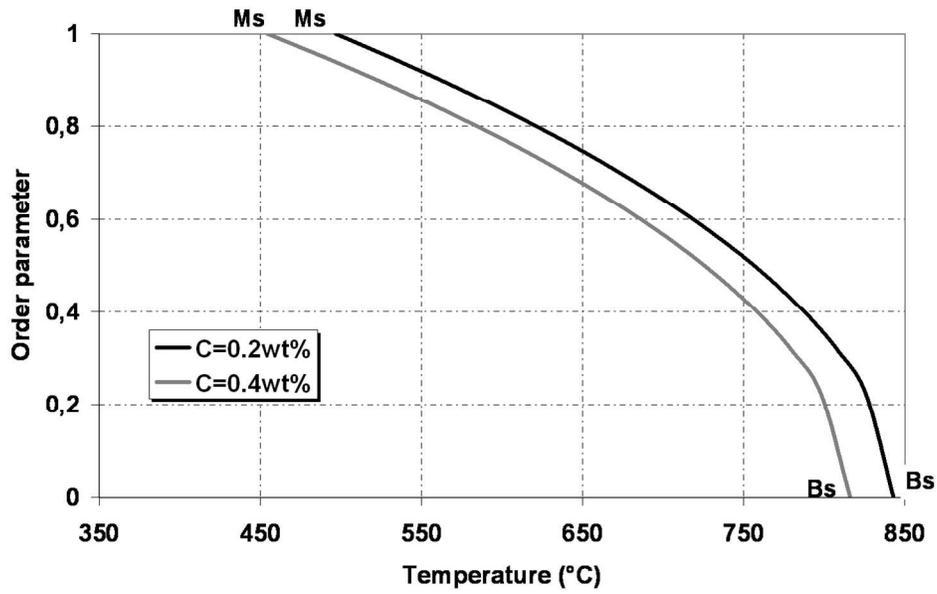

256x159mm (150 x 150 DPI)

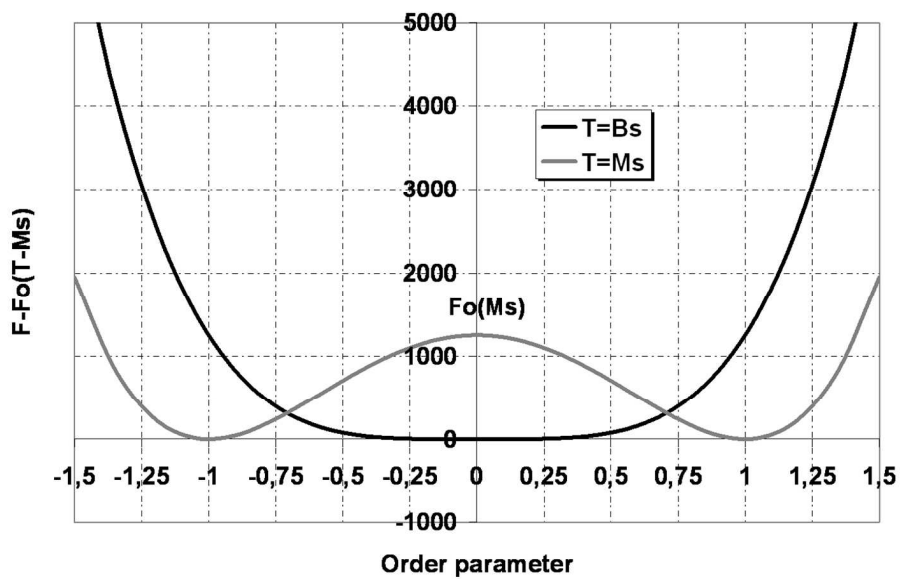

256x159mm (150 x 150 DPI)

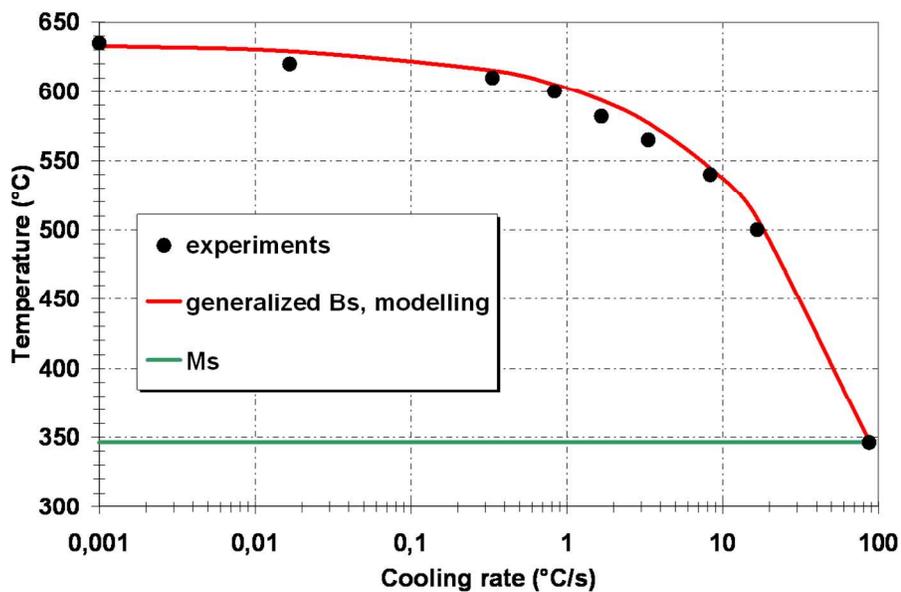

256x159mm (150 x 150 DPI)